\documentclass[a4paper,11pt]{article}
\usepackage{pos}

\usepackage{graphicx}  % needed for figures
\usepackage{dcolumn}   % needed for some tables
\usepackage{bm}        % for math
\usepackage{amsmath}
\usepackage{epsfig}
\usepackage{rotating}
\usepackage{xspace}
\usepackage{hyperref}
\usepackage{xcolor}
\newcommand{\C}{\mathbb{C}}

\newcommand{\R}{\mathbb{R}}

\newcommand{\cO}{{\cal O}}

\newcommand{\real}{{\rm Re}}
\newcommand{\imag}{{\rm Im}}

\newcommand{\bra}{\langle}
\newcommand{\ket}{\rangle}

\newcommand{\bear}{\begin{eqnarray}}
\newcommand{\eear}{\end{eqnarray}}

\def\be#1\ee{\begin{equation}#1\end{equation}}
\def\bea#1\eea{\begin{align}#1\end{align}}

\title{Complex Langevin: Boundary terms at poles of the drift}
\author*[a]{Erhard Seiler}
\affiliation[a]{Max-Planck-Institut f\"ur Physik (Werner-Heisenberg-Institut),\\
  F\"ohringer Ring 6, M{\"u}nchen, Germany}

\emailAdd{{ehs@mpp.mpg.de}}
\abstract
{The complex Langevin method is a general method to treat systems with complex 
action, such as QCD at nonzero density. The formal justification relies on the 
absence of certain boundary terms, both at infinity and at the unavoidable poles 
of the drift force. Here I focus on the boundary terms at these poles for simple 
models, which so far have not been discussed in detail. The main result is that 
those boundary terms (for the ``un-evolved'' observables) arise after running 
the Langevin process for a finite time and vanish again as the Langevin time 
goes to infinity. This is in contrast to the boundary terms at infinity, which 
can be found to occur in the long time limit (cf.~the contribution by D\'enes 
Sexty).
}

\FullConference{%
 The 38th International Symposium on Lattice Field Theory, LATTICE2021
  26th-30th July, 2021
  Zoom/Gather@Massachusetts Institute of Technology
}

\begin{document}
\maketitle

\section{Introduction} 

This contribution is based largely on \cite{seiler}, to which we refer for
more details.

To be able use stochastic sampling methods for a complex holomorphic density 
$\rho\propto e^{-S}$ on ${\R^d}$, one searches for a probability density 
${P\ge 0}$ on ${\C^d}$, s.t.

\be
\bra \cO\ket \equiv \int_{\R^d} \cO(x) \rho(x) dx=
\int_{\C^d} \cO(x+iy) P(x,y) dx dy\,.
\ee
for {\em holomorphic} observables ${\cO}$.

Klauder \cite{klauder} and Parisi \cite{parisi} proposed a general, very 
flexible method to produce such a $P$, as the equilibrium distribution of 
{\em real stochastic process} on ${\C^d}$, called {\em Complex Langevin} 
(CL) method, by defining
\be 
dz=K dt+dw, \quad K=-\nabla S 
\label{cprocess}
\ee 
($dw$ is the {real} Wiener increment normalized as $\bra dw^2\ket=2$). 
This process necessarily wanders into the complex realm ${\C^d}$. Written 
out (\ref{cprocess}) becomes
\be
dx=K_x dt+ dw,\quad K_x=\real\,K
\ee
\be
dy=K_y dt,\quad\quad\quad\;\, K_y=\imag\,K
\ee
The hope is then that the long time average of this process yields the 
$\rho$-expectation. A strategy to justify this was proposed in \cite{trust}. 
Briefly it proceeds as follows: we want for a sufficiently large class of 
observables
 $\cO$
\be
\bra \cO\ket_{\rho(t)}=\bra \cO\ket_{P(t)} \quad\forall\,t\ge 0\,,
\label{want}
\ee
where $P(t)\equiv P(x,y;t)$ is the probability density of the CL stochastic 
process and $\rho(t)\equiv \rho(x;t)$ is the evolved complex density solving 
the ``complex Fokker-Planck equation''
\be
\frac{\partial}{\partial t}\rho(z;t)= L_c^T \rho(z;t)\,,\quad 
L_c^T=\partial_z(\partial_z-K(z))\,.
\label{complexfp}
\ee
with initial conditions such that (\ref{want}) holds at $t=0$.
$L_c^T$ is the transpose of $L_c$, governing the evolution of observables:  
\be
\frac{\partial}{\partial t}\cO(z;t)=L_c \cO(z;t)\,,\quad
L_c=(\partial_z+K(z))\partial_z\,.
\ee
It is easy to see that
\be
\int dx \rho(x;t)\cO(x;0)=\int dx \rho(x;0)\cO(x;t)
\ee
provided the integration connects two zeroes (finite or infinite) of $\rho$.
Eq.(\ref{want}) is then true if the function
\be
{F_\cO(t,\tau)\equiv \int P(x,y;t-\tau) \cO(x+iy;\tau)dx\,dy}
\quad (0\le \tau\le t)
\ee
is independent of $\tau$, i.e.
\be
\frac{\partial}{\partial\tau} F_\cO(t,\tau)=0\quad\forall t\ge 0\,.
\label{boundterm}
\ee
Here $\cO(z;\tau)$ is the solution of the initial value problem
\be
\frac{\partial}{\partial \tau}\cO(z;\tau)=L_c\cO(z;\tau)\,,\quad
\cO(z;0)=\cO(z)\,;\quad L_c=(\partial_z+K(z))\partial_z\,,
\label{cauchy}
\ee

(\ref{want}) follows from (\ref{boundterm}) because $F_\cO(t,\tau)$ 
interpolates between two sides of (\ref{want}), as shown in \cite{trust}
(assuming integration by parts in $x$ without boundary terms).
 Integration by parts in $x,y$ shows that (\ref{boundterm}) holds, up to 
possible boundary terms, in other words $\frac{\partial}{\partial\tau} 
F(t,\tau)$ {\bf is} a (sum of) boundary terms. Boundary terms may arise 
at infinity as well as at poles.
 
An important caveat that was stated in \cite{trust} is the following: 
(\ref{want}) implies correctness of CL only if
\be
\lim_{t\to\infty} \bra \cO\ket_{\rho(t)}
\ee
exists and is unique. i.~e. if the spectrum of $L_c^T$ lies in the left half 
of $\C$ and $0$ is a simple eigenvalue with eigenfunction $\rho(x)$ (see the 
remark at the end of Section 3). 

Possible failure of the CL method was analyzed from a different point of view by 
Salcedo \cite{salc}; the problem caused by poles of the drift was studied by 
Nishimura and Shimasaki in simple models \cite{nishi}; in \cite{pole} we 
presented a detailed study of this issue, with the emphasis on numerical 
analysis of various models, from the simplest one-dimensional case to full QCD. 
Here we trace the problem to the occurrence of boundary terms.

\section{The need to consider the evolution before reaching equilibrium}
\label{finite}

In \cite {scherzer1,scherzer2} we found boundary terms at infinity by 
considering the equilibrium distributions and ``un-evolved'' observables, 
i.~e. $\partial_\tau F(t,\tau)|_{\tau=0}$  (cf. D.~Sexty, these 
proceedings).  

But this type of boundary term does not appear at poles. This is because 
empirically the equilibrium distribution $P(x,y;t=\infty)$ vanishes at 
least linearly at the poles of the drift, so holomorphic observables 
could not lead to boundary terms there. (Note that this argument does not hold for
``evolved'' observables, which are at best meromorphic.)

To see this in a little more detail, let's consider for simplicity a pole at the 
origin; consider the approximate boundary term
\be
\int_{x^2+y^2\le\delta^2} dx\,dy P(x,y;t=\infty) 
L_c\cO(x+iy)\,.
\label{volume}
\ee
Using the Cauchy-Riemann equations and integrating by parts (\ref{volume}) 
becomes
\be 
\int_{x^2+y^2\le\delta^2}  dx\,dy \cO(x+iy) (L^TP)(x,y;t=\infty) 
+B_{\delta}= B_{\delta}\,
\label{boundterm2}
\ee
where $L^T$ is the {\em real} Fokker-Planck operator 
\be
L^T=\partial_x^2-\partial_xK_x -\partial_y K_y
\ee
describing the evolution of $P$ under the stochastic process, see 
\cite{trust}). (\ref{boundterm2}) holds since the first term of the 
left-hand side vanishes in equilibrium. $B_{\delta}$ is a boundary term. 
Now, since $\cO$ is holomorphic, $L_c\cO$ has at most a simple pole at 
the origin, stemming from the pole in the drift. Since $P$ vanishes 
linearly at the origin, the integrand of (\ref{volume}) is bounded in 
the region of integration, hence the boundary term vanishes for 
$\delta\to 0$. 

If instead we consider the time evolution for finite time $t$, 
$B_\delta$ now is given by
\be
B_\delta=\int_{x^2+y^2\le\delta^2} dx\,dy \left\{\cO(x+iy)L^T 
P_{z_0}(x,y;t)-P_{z_0}(x,y;t) L_c\cO(x+iy)\right\}\,
\label{full_bt}
\ee
and the first term of this expression is no longer zero. Since the 
equilibrium distribution does not lead to a boundary term, we now 
consider the evolution for short times.

\section{One-pole model}

The one-pole model is defined by
\be
\rho(x)= (x-z_p)^{n_p} \exp(-\beta x^2)\,
\ee
with $n_p$ a positive integer.

\begin{figure}[ht]
\begin{center}
\includegraphics[width=0.48\columnwidth]{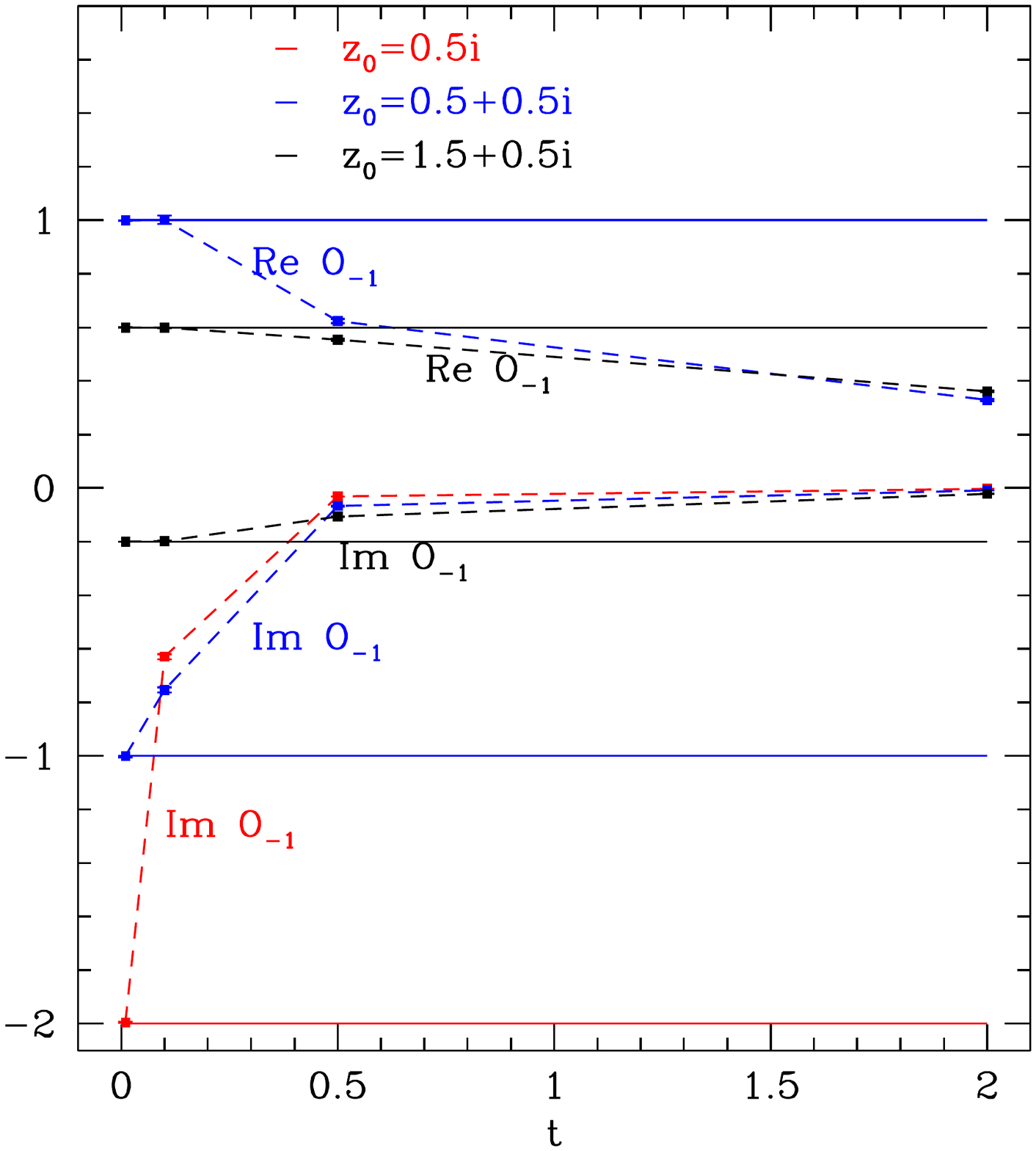}
\includegraphics[width=0.48\columnwidth]{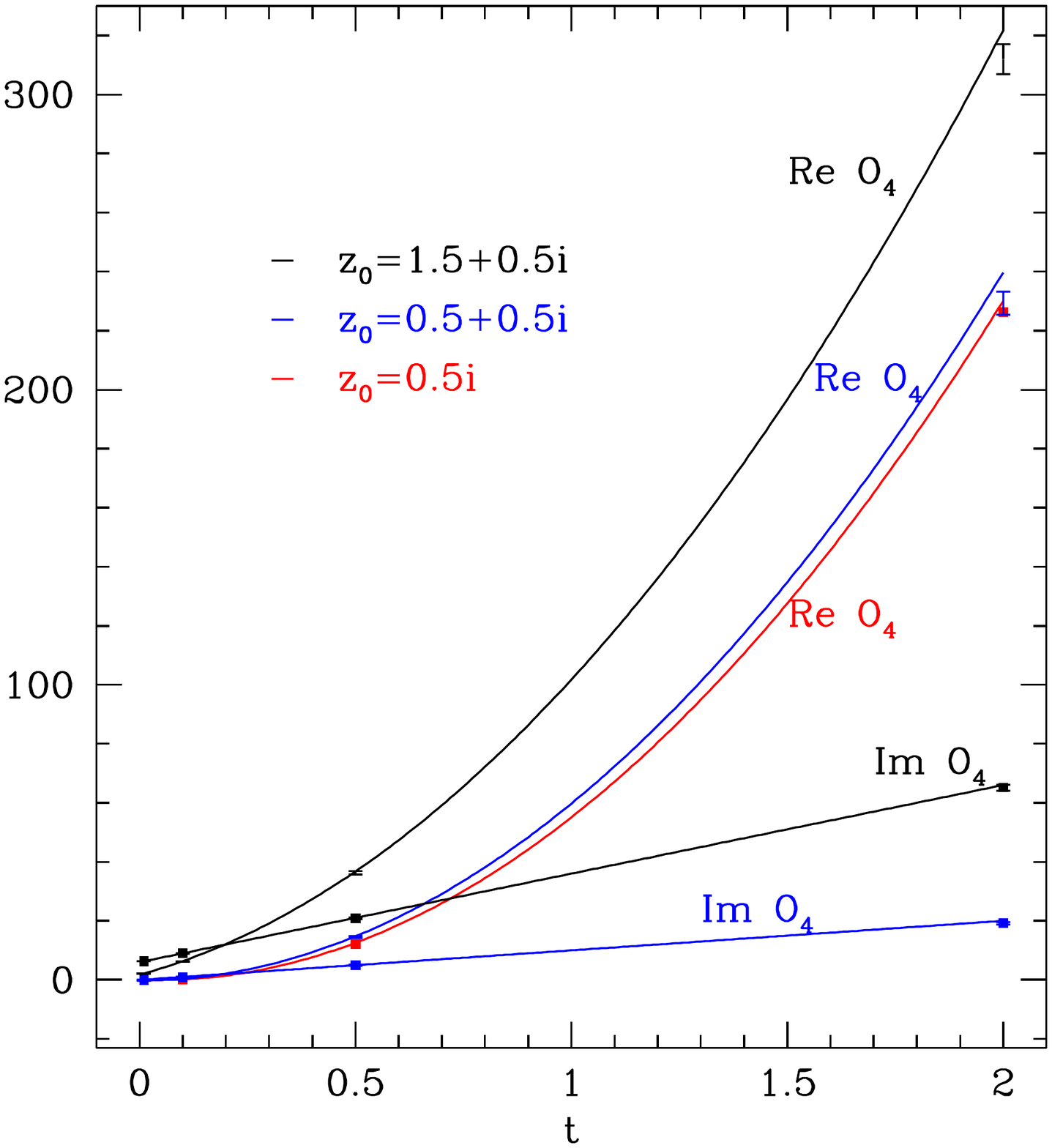}
\vglue-20mm
\caption{Pure pole model. Comparison of two evolutions for $k=-1$ (left) and $k=4$ 
(right). Dashed lines: connecting CL results, solid lines: analytic $\rho$ evolution 
(\ref{ppevol}).}
\label{fpecomp}
\end{center}
\end{figure}

Since we are not interested in large times, we can simplify the model even 
further by putting $\beta=0$, giving rise to the ``pure pole model''; without
loss we also set $z_p=0$. For the special case $n_p=2$ there is an explicit 
formula for the integral kernel of $\exp(tL_c)$:
\be
\exp(tL_c)(z,z')
=\frac{z'}{z\sqrt{4\pi t}}\exp\left(\frac{(z-z')^2}{4t}\right)\,,
\label{ppkernel}
\ee
where $z'=x'+iy_0$ and the integration is over $x'$. As observables 
we take the powers
\be
\cO_k(z)\equiv z^k\,, k=-1,0,1,2\ldots\,.
\ee
Using (\ref{ppkernel}) we can explicitly compute the evolution of those observables,
obtaining e.~g.
\bea
&\cO_{-1}(z;t)=\frac{1}{z}\,.\notag\\
&\cO_1(z;t)= z+\frac{2t}{z}\,,\notag\\
&\cO_2(z;t)=z^2+6t\,,\notag\\
&\cO_3(z;t)=z^3+12 t z+\frac{12t^2}{z}\,,\notag\\
&\cO_4(z;t)=z^4+20 t z^2+60t^2\,,
\label{ppevol}
\eea
The fact that no higher negative powers occur for $n_p=$ was already noted in 
\cite{pole}. We compared these results with those of the CL evolution 
$\bra\cO_k\ket_{P(t)}$ obtained by running $10^5$ CL trajectories, all with the same 
starting point $z=z_0$ up to the desired time $t$. The comparison is shown in 
Fig.~\ref{fpecomp} for $k=-1$ and $k=4$.

As in these examples, generally for even powers there is agreement, 
whereas for odd powers already at small times there is disagreement, 
signaling the presence of boundary terms for these powers.

As long as $n_p=2$ we also have a closed expression for the integral kernel 
$\exp(tL_c)$ for $\beta>0\,,z_p=0
\,,n_p=2$:
\bea
\exp(tL_c)(z,z')
=& \frac{z'}{z}\exp\left(\frac{\beta}{2}(z^2-z'^2)\right)\exp(2\beta t)
\notag\\
\times\sqrt{\frac{\beta}{\pi (1-e^{-4\beta t})}}
&\exp\left[-\frac{\beta (z^2+z'^2)}{2\tanh(2\beta t)}\right]
\exp\left(\frac{\beta z z'}{\sinh(2\beta t)}\right)\,
\label{mehler}
\eea
(based on Mehler's formula \cite{barry}). Here again $z'=x'+y_0$,\; $x'$ being 
the integration variable.

Defining
\be
b\equiv\frac{\beta}{\sinh(2\beta t)}\,;\quad
\sigma\equiv\frac{1-\exp(-4\beta t)}{2\beta}\,,
\ee
we find for the same observables as above 
{{\begin{align}
&\cO_2(z;t)=3\sigma+b^2\sigma^2 z^2\to \frac{3}{2\beta}
\quad (t\to\infty)\notag\\
&\cO_4(z;t)=15\sigma^2+ 10 b^2\sigma^3 z^2+b^4\sigma^4 z^4
\to \frac {15}{4\beta^2} \quad (t\to\infty)
\notag\\
&\cO_1(z;t)=\frac{1}{bz}+b\sigma \to\infty  \quad (t\to\infty)
\notag\\
&\cO_3(z;t)=\frac{3\sigma}{bz}+6 b\sigma^2 z
+b^3\sigma^3 z^3 \to\infty  \quad (t\to\infty)
\notag\\
&\cO_{-1}(z;t)=\frac{1}{b\sigma z}\to\infty  \quad (t\to\infty)
\label{betaevol}
\end{align}}
Even powers remain bounded for $t\to\infty$ and actually converge to the
correct limit, whereas odd powers grow exponentially!

This signals the presence of an exponentially growing mode in both 
$\exp(tL_c)$ and its transpose $\exp(tL_c^T)$, so in this case $\bra 
\cO\ket_{\rho(t)}$ is {\emph not} the correct evolution. This failure of 
correctness is again due to the existence of boundary terms: absence of 
boundary terms also implies absence of exponentially growing modes 
as will be demonstrated elsewhere  \cite{seiler2}.

\section{Direct numerical evaluation of the boundary term} 

We now consider $\beta>0, z_P\neq 0$. In this case we do not have an 
analytic expression for the integral kernel of $\exp(ptL_c)$, but we can 
numerically evaluate the approximate boundary term $B_\delta$ (\ref{full_bt}). 

After some easy manipulations we find for {\em any} holomorphic observable 
$\cO(z)$ 
\be
B_\delta= -\oint_{|z-z_p|=\delta} \vec K \cdot \vec n \,
P_{z_0}(x,y;t)\cO(x+iy)ds+o(\delta)=  
-\oint_{|z-z_p|=\delta} \vec K \cdot \vec n P_{z_0}(x,y;t)\cO(0)ds+o(\delta) \,,
\label{circle}
\ee
We approximate the circle of radius $\delta$ in (\ref{circle}) by a thin ring of 
thickness $\eta\delta$:
\be
(1-\eta) \delta <|z-z_p|<(1+\eta)\delta\,.
\ee
For $n_p=2,\beta=1 ,z_p=-i/2, z_0=i/2$ and $\cO(0)=1$ the CL process gives the 
results shown in Fig.~\ref{apprbc}.

\begin{figure}[ht] 
\begin{center}
\includegraphics[width=0.48\columnwidth]{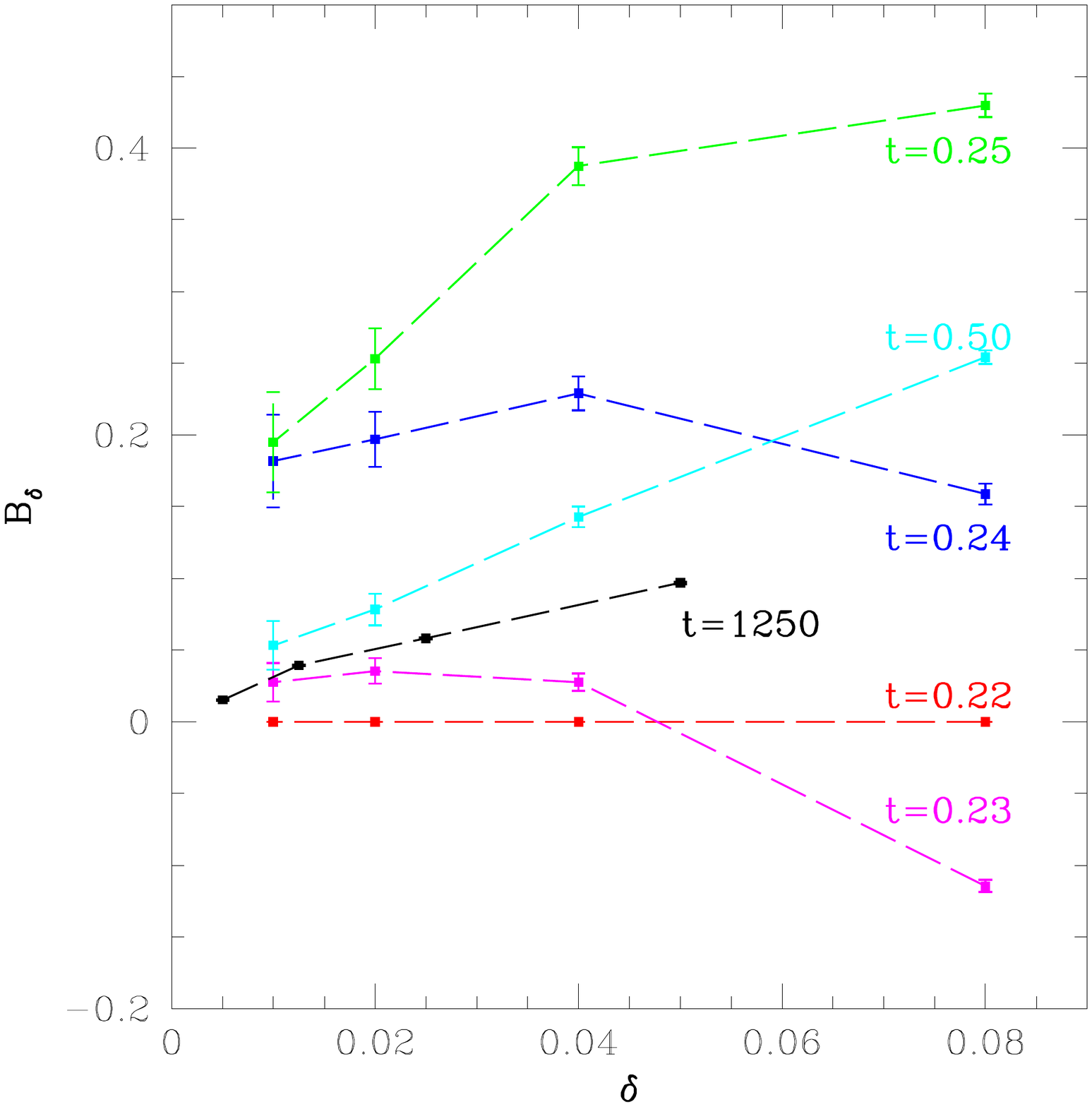}
\includegraphics[width=0.48\columnwidth]{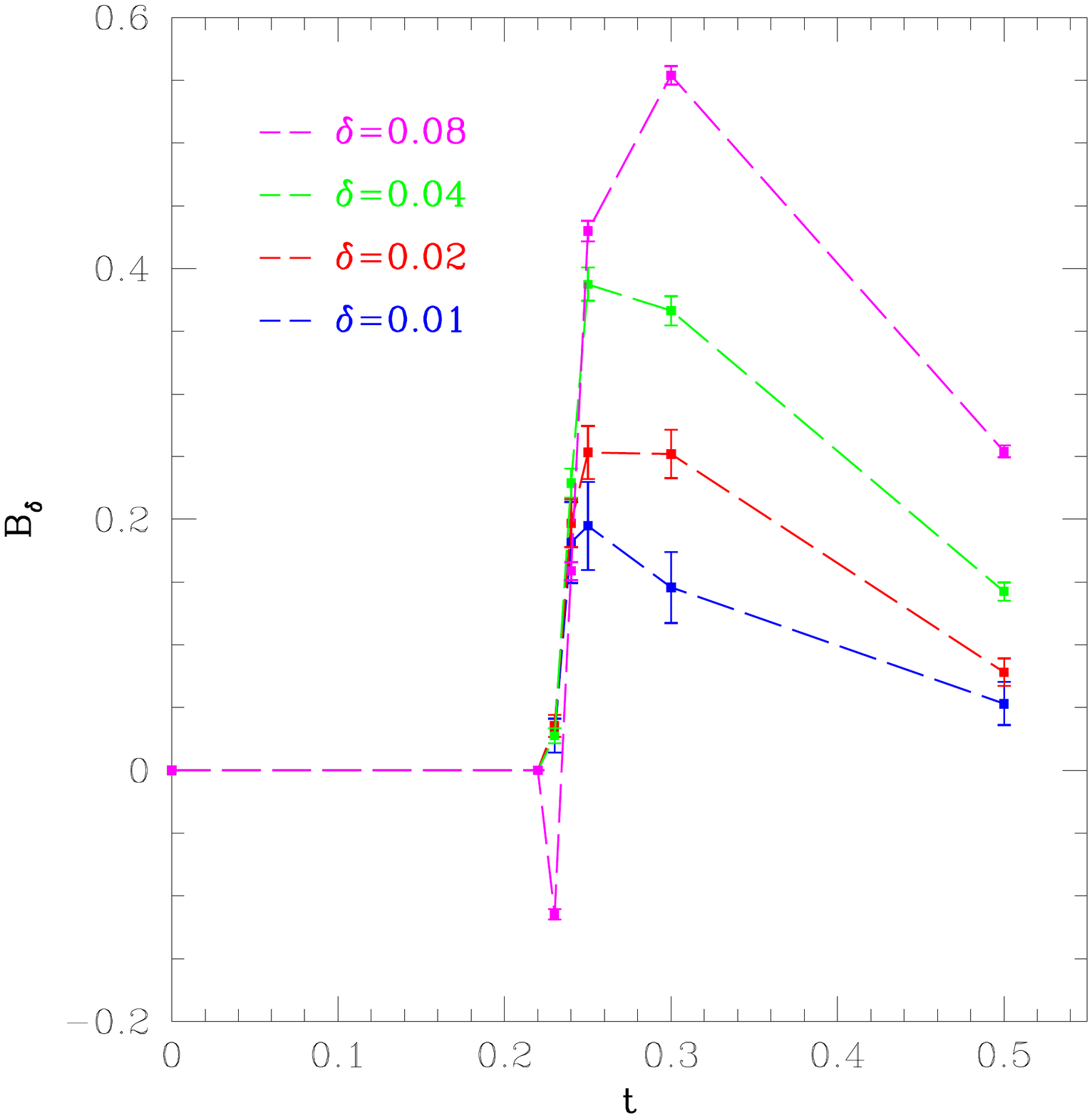}
\vglue-2cm
\caption{Full pole model. Numerical estimates of the boundary term $B_\delta$ for 
$\eta=0.1$\;. Left: $B_\delta$ vs. $\delta$, right: $B_\delta$ vs. $t$.}
\label{apprbc}
c\end{center}
\end{figure}

The CL data are produced as before by running trajectories up to the respective 
times $t$. Because the chance of hitting the small rings is so small, we took 
here $10^7$ trajectories, but we still obtained only about 50 hits for 
$\delta=0.01,\, \eta=0.1$ and $t=0.24$ and $0.25$, and even fewer for other 
values of $t$. Therefore the extrapolation to $\delta=0$ can only be to be done 
``by eye''. But Fig.~\ref{apprbc} still shows clearly that for $t<0.22$ there is 
no boundary term. This is because the process takes at least that much time to 
move from the starting point to the location of the pole. (Because there is no 
noise in the $y$ direction one can compute this time by evaluating a simple 
integral.)

For $0.23< t \le 0.25$ there are clear indications of a boundary term, whereas for 
$t>0.5$ it is fading away, and it has disappeared for $t=1250$, where we get much 
smaller statistical errors because we can thermalize and average over initial 
conditions.

But it should be stressed that we were considering $\partial_\tau F(t,\tau)$ 
only for $\tau=0$. It is to be expected that for $t>0.5$ the boundary term 
reappears at nonzero values of $\tau$.
\vskip3mm  
I would like to thank all the people with whom I had the pleasure of collaborating
on the CL method: Gert Aarts, Felipe Attanasio, Lorenzo Bongiovanni, Pietro Giudice,
Benjamin J\"ager, Frank James, Jan Pawlowski, Lorenzo Luis Salcedo,
Manuel Scherzer, D{\'e}nes Sexty, Ion-Olimpiu Stamatescu, Jacek Wosiek.

\end{document}